\documentclass[11pt,twoside]{article}

\usepackage{asp2006_hotcool}
\usepackage{epsf}
\usepackage{graphicx}

\begin{document}
\title{Radiative Transfer Modeling of the Winds \& Circumstellar Environments of Hot And Cool Massive Stars}  
\author{A. Lobel}  
\affil{Royal Observatory of Belgium, Ringlaan 3, B-1180 Brussels, Belgium}   

\begin{abstract}
We present modeling research work of the winds and circumstellar environments of a variety of prototypical hot and cool massive stars using advanced radiative transfer calculations. This research aims at unraveling the detailed physics of various mass-loss mechanisms of luminous stars in the upper portion of the H-R diagram. 
Very recent 3-D radiative transfer calculations, combined with hydrodynamic simulations, show that radiatively-driven winds of OB supergiants are structured due to large-scale density- and velocity-fields caused by rotating bright spots at the stellar equator. The mass-loss rates computed from matching Discrete Absorption Components (DACs) in $IUE$ observations of HD 64760 (B Ib) do not reveal appreciable changes from the rates of unstructured (smooth) wind models. 

Intermediate yellow supergiants (such as the yellow hypergiant $\rho$ Cas, F$-$G Ia0), on the other hand, show prominent spectroscopic signatures of strongly increased mass-loss rates during episodic outbursts that cause dramatic changes of the stellar photospheric conditions. Long-term high-resolution spectroscopic monitoring of cool hypergiants near the Yellow Evolutionary Void reveals that their mass-loss rates and wind-structure are dominated by photospheric eruptions and large-amplitude pulsations that impart mechanical momentum to the circumstellar environment by propagating acoustic (shock) waves. 

In massive red supergiants, however, clear evidence for mechanical wave propagation from the sub-photospheric convection zones is lacking, despite their frequently observed spectroscopic and photometric variability. Recent spatially resolved HST-$STIS$ observations inside Betelgeuse's (M Iab) very extended chromosphere and dust envelope show evidence of warm chromospheric gas far beyond the dust condensation radius of radiative transfer models. Models for these long-term spectroscopic observations demonstrate that the chromospheric pulsations are not spherically symmetric. The $STIS$ observations point to the importance of mechanical wave propagation for heating and sustaining chromospheric conditions in the extended winds of red supergiants.
\end{abstract}

\section{Introduction}

The winds and circumstellar environments of massive stars 
are unique laboratories for studying the detailed physics of 
stellar mass-loss mechanisms. In many luminous hot and cool 
supergiants large spectroscopic variability is observed 
on time-scales of weeks to years. Long-term spectroscopic 
monitoring of massive-star variability over the past three 
decades therefore has provided a wealth of information
about the driving physics and active structuring mechanisms
in these stars with very extended winds. This paper provides a brief 
overview of important research results for a hot, intermediate, 
and cool supergiant obtained with advanced radiative transfer modeling 
of the spectroscopic variability. 

\section{3-D Modeling the Wind of Hot Supergiant HD~64760 (B Ib)}

In recent years it has become clear that mass-loss rates of hot 
massive stars are considerably overestimated because the clumping 
of their winds has not been taken into account. Understanding the physical 
nature of these wind clumping processes, determining the amount of wind 
clumping, and revising the total stellar mass-loss rates are therefore 
hot topics in current massive star research. Detailed information 
about the wind structure can be obtained with accurate numerical models. 
We develop a parallel computer code {\sc Wind3D} which solves 
the 3-D transport of radiation in the scattering winds of massive stars. 
It calculates the detailed shape of resonance lines that form in 
very extended winds (e.g., P Cygni profiles) in non-LTE.  

\begin{figure*}[!t]
\begin{center}
\includegraphics
 [width=379pt,height=167pt]{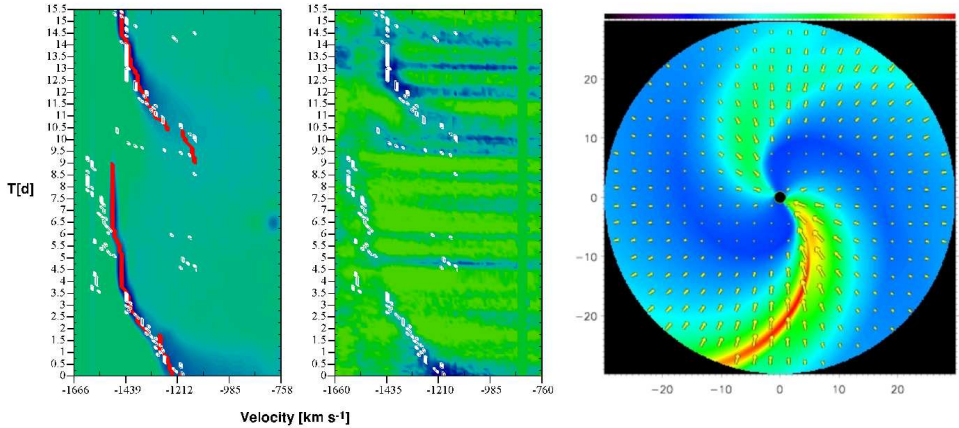}
\caption{{\it Panel left:} 3-D non-LTE radiative transfer fit 
with {\sc Wind3D} to the DACs in Si~{\sc iv} $\lambda$1395 
compared to $IUE$ observations of HD 64760 ({\it middle panel}). 
{\it Panel right:} 
Density contrast of the hydrodynamic input model computed with {\sc Zeus3D}.
Two unequally bright spots produce large-scale wind-velocity structures 
and density enhancements. The size of the over-plotted arrows 
indicates the magnitude of the velocity deceleration with respect to 
the smooth unperturbed wind. The bright spots cause rotating 
density waves in the equatorial wind of the hot supergiant ({\it see text}).
\label{lobel:fig1}}
\end{center}
\end{figure*}

We apply {\sc Wind3D} for the detailed modeling of DACs observed in the
Si~{\sc iv} resonance lines of HD 64760 (B0.5 Ib), a key massive hot star. 
The study provides new evidence that the DACs are in fact large-scale 
spiraling density- and velocity-structures in the equatorial wind. 
These wind structures are due to rotating hotspots at the stellar 
surface. The bright spots do not co-rotate with the stellar surface, 
but lag five times behind the fast surface rotation.
A hydrodynamic model with two spots of unequal brightness and size
on opposite sides of the equator, with opening angles of 
20$\deg$ $\pm$5$\deg$~and 30$\deg$ $\pm$5$\deg$~diameter, 
and that are 20$\pm$5\% and 8$\pm$5\% brighter than the 
stellar surface, respectively, provides the best fit to the 
observed DACs ({\it Fig. 1}). The mass-loss rate of the 
structured wind model does not exceed the rate of the 
spherically symmetric smooth wind model by more than 1\% 
\citep{lob08}.

\begin{figure*}[!t]
 \plottwo{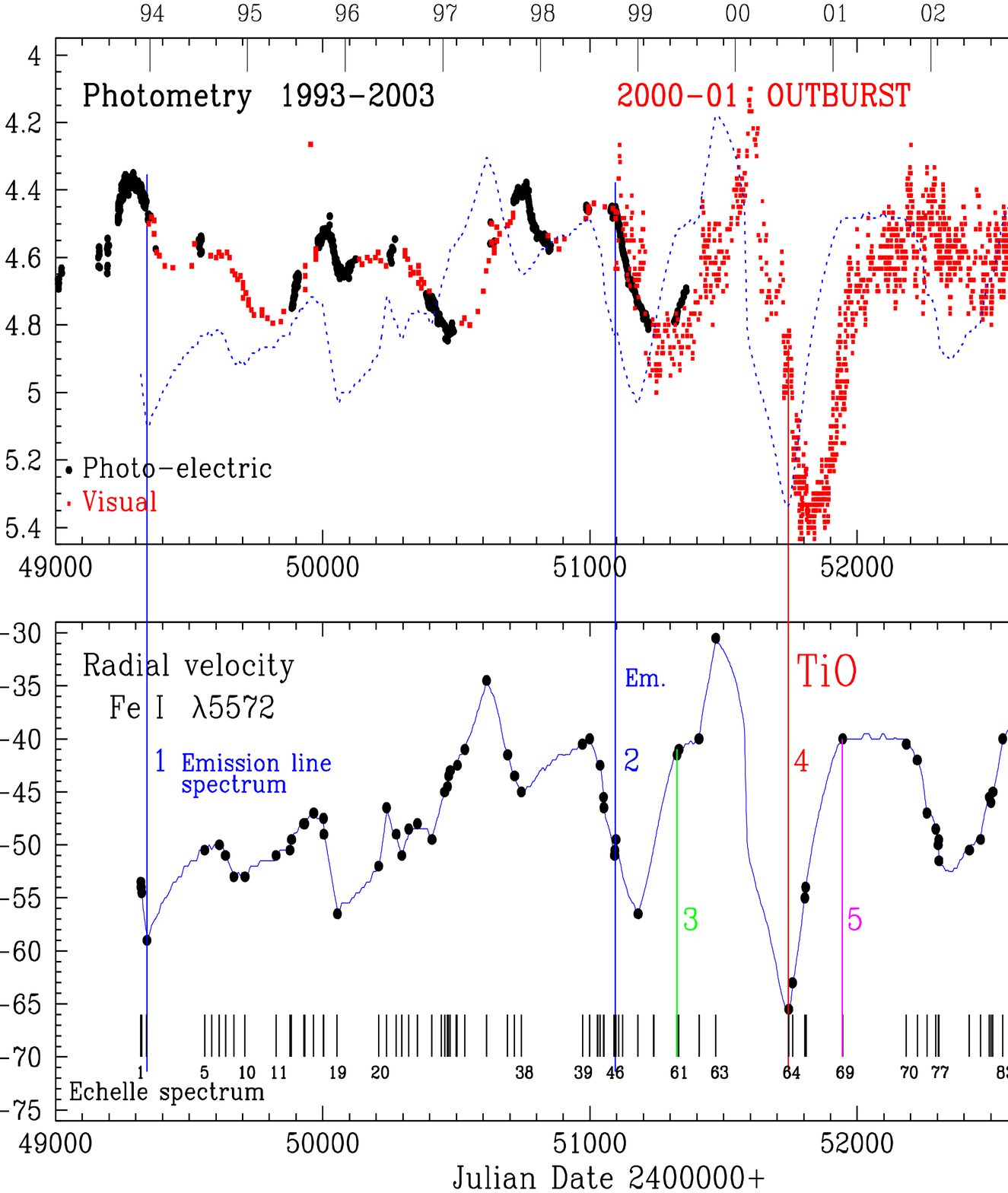}{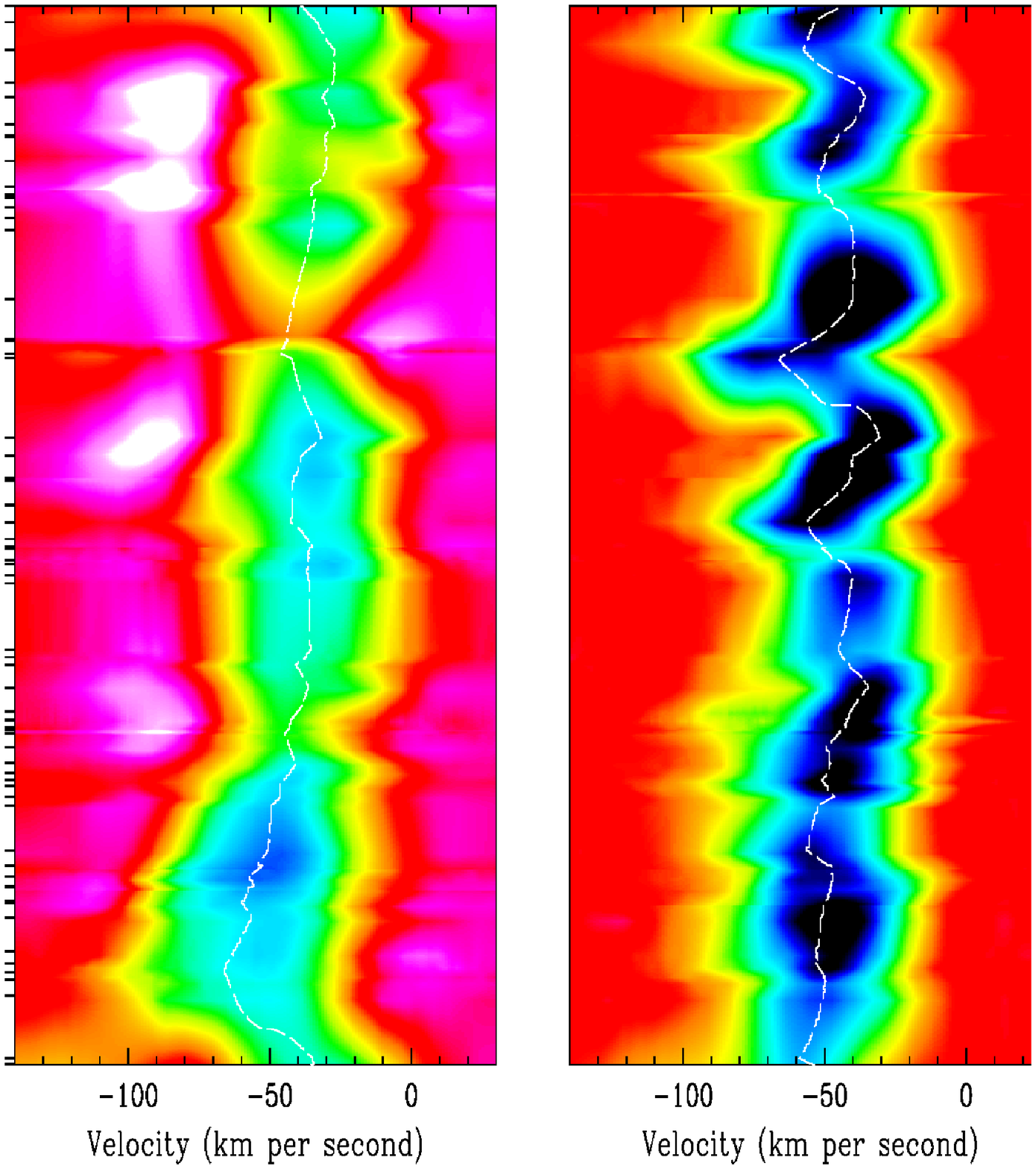} 
\caption[]{{\it Panel left:} Visual brightness changes observed 
in $\rho$ Cas between 1993 and 2003 ({\it upper panel}) are compared 
to the radial velocity variability ({\it lower panel}) of an 
unblended photospheric absorption line of Fe~{\sc i} $\lambda$5572.
{\it Panel right:} Dynamic spectra of H$\alpha$ ({\it left}) and Fe~{\sc i}
$\lambda$5572 ({\it right}). Observations are marked with 
left-hand tickmarks. The Fe~{\sc i} line strongly blue-shifts during 
the large outburst of mid 2000 when the atmosphere cools by 
3000 K ({\it see text}).
\label{lobel:fig2}}
\end{figure*}

\section{The Millennium Outburst of Yellow Hypergiant $\rho$ Cas (F$-$G Ia0)}

Yellow hypergiants are post-red supergiants, rapidly evolving toward 
the blue supergiant phase \citep{dej98}. They are among the prime candidates 
for progenitors of Type II supernovae in the Galaxy.
In July 2000 we observed the formation of new TiO bands in F-type hypergiant
$\rho$ Cas during a strong $V$-brightness decrease of $\sim$$1^{\rm m}.4$ 
({\it Fig. 2}). Synthetic spectrum calculations reveal that 
$T_{\rm eff}$ decreased by at least 3000 K, from 7250 K to $\simeq$4250~K, 
and the spectrum became comparable to an early M-type supergiant. 
The TiO bands signal the formation of a cool circumstellar gas shell 
with $T_{\rm gas}$$<4000$ K due to the supersonic expansion of the photosphere 
and upper atmosphere. We observe a shell expansion velocity of 
$\sim$35$\pm$2~$\rm km\,s^{-1}$ from the TiO bands. From the 
synthetic spectrum fits to the bands we compute an exceptionally 
large mass-loss rate of 
$\dot{M}$$\simeq$5.4\,$\times$\,$10^{-2}$~$\rm M_{\odot}\,yr^{-1}$,
($\sim$5\% of $\rm M_{\odot}$ in 200 d). It is therefore one of the largest stellar 
outburst events that has directly been monitored so far \citep{lob03}.
The right-hand panel of Fig. 2 shows dynamic spectra of H$\alpha$ 
and Fe~{\sc i} $\lambda$5572 observed between 1993 and late 2003. 
The radial velocity curves of the H$\alpha$ absorption core and the 
photospheric Fe~{\sc i} lines ({\it white dashed lines}) reveal a velocity 
stratified dynamic atmosphere. A strong collapse of the upper 
H$\alpha$ atmosphere and the lower photosphere precedes the outburst 
event during the large-amplitude pre-outburst pulsation cycle of 1999.

Based on radiative transport calculations \citet{gor06} presented 
a model for the near-IR CO emission emerging from cool atmospheric layers 
in the immediate vicinity of $\rho$ Cas' photosphere. The gas kinetic 
temperature minimum in the model results from a periodical pulsation-driven 
shock wave. 

\begin{figure*}[!t]
\plottwo{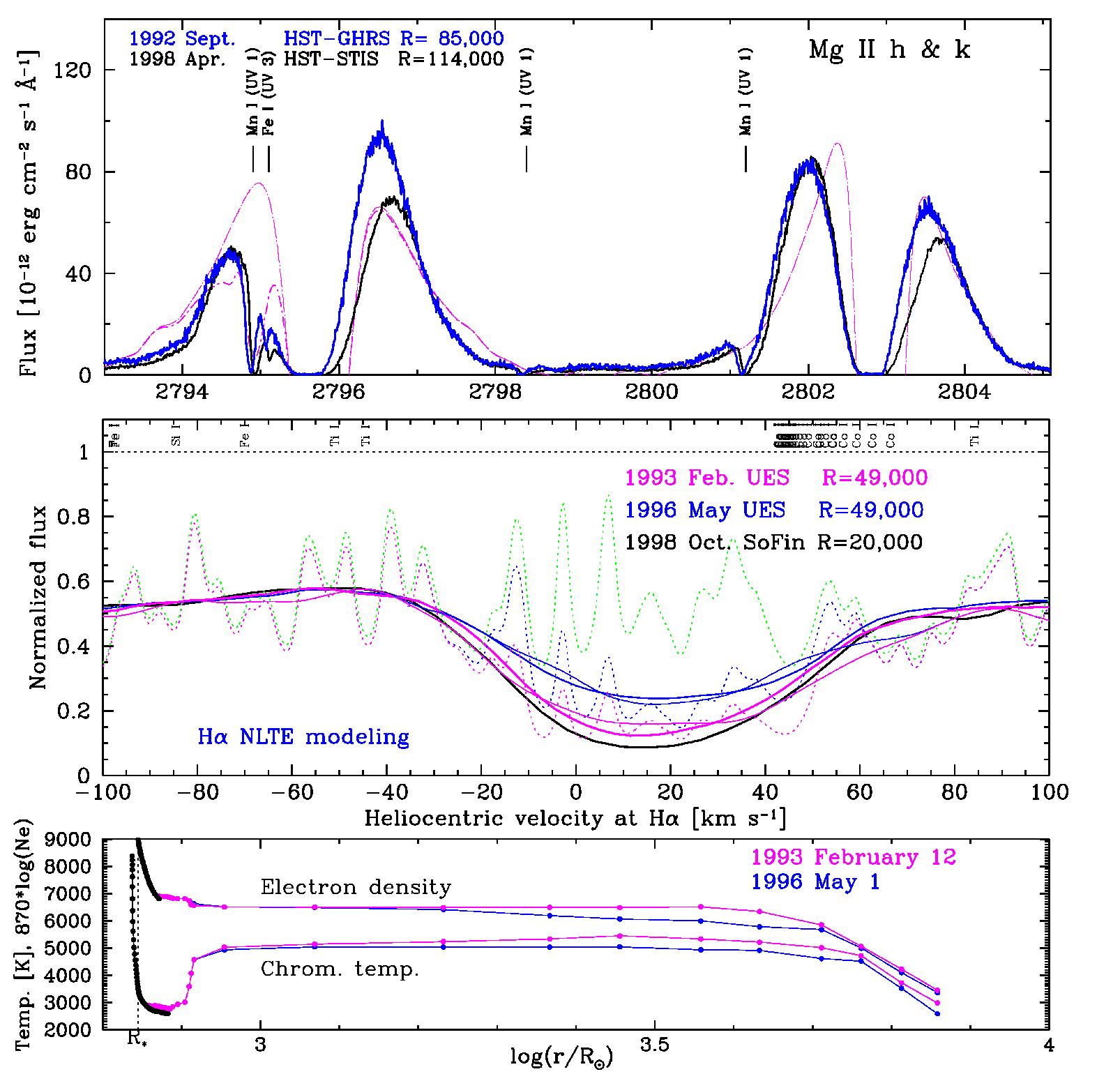}{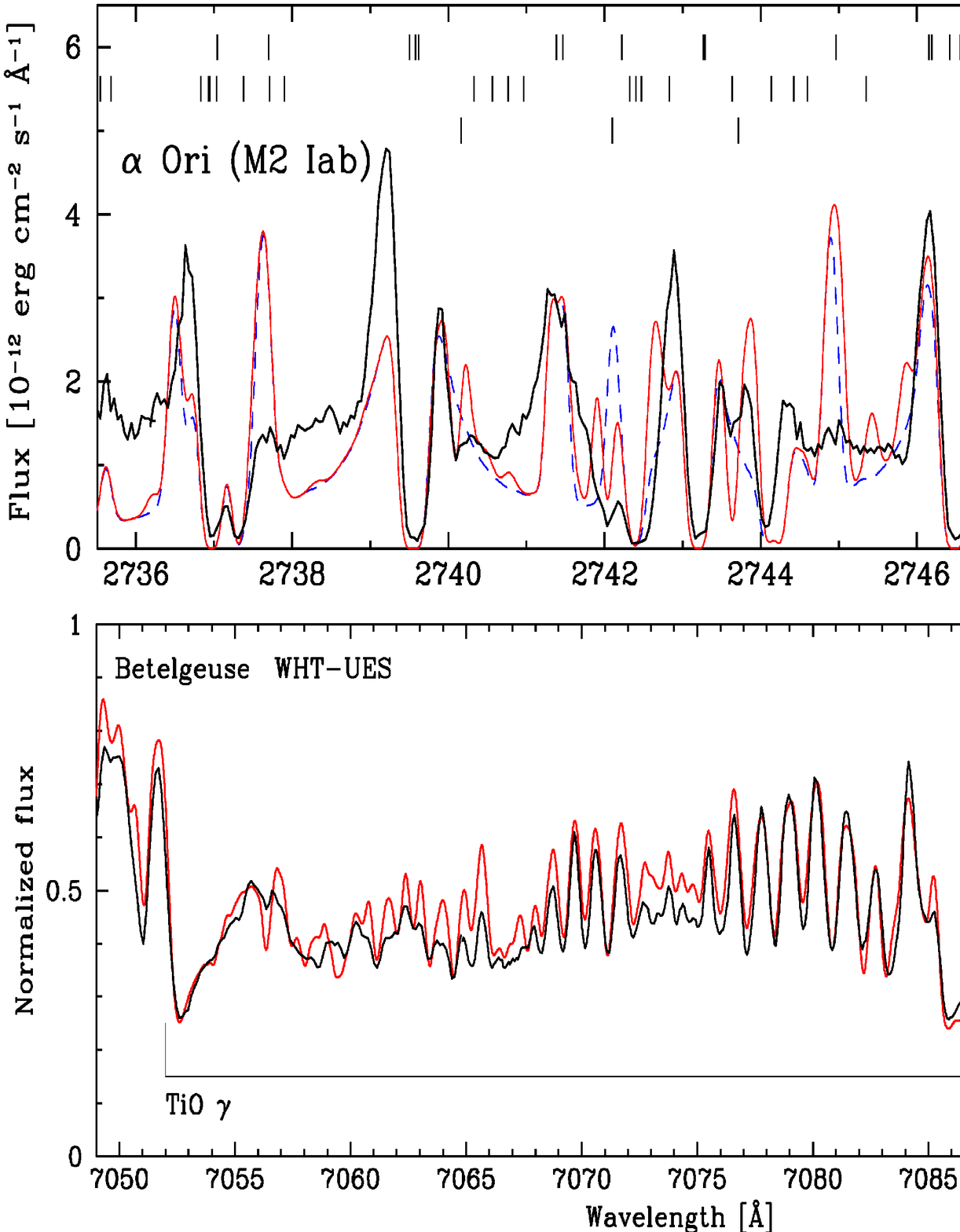} 
\caption[]{
{\it Panels left:} Large spectral resolution observations of the 
Mg~{\sc ii} $h$ \& $k$ and Balmer H$\alpha$ lines 
({\it boldly drawn lines}) reveal long-term chromospheric 
variability in $\alpha$ Ori. Best non-LTE fits 
({\it thin drawn lines}) constrain the gas kinetic temperature 
and electron density structure of the inner chromosphere. The model
of the photosphere below the chromosphere has $T_{\rm eff}$=3500 K.
({\it bottom panel}). 
{\it Panels right:}  
The chromospheric emission line spectrum ({\it upper panel}) is 
well-reproduced using only iron lines ({\it dashed drawn line}), 
and a model $T_{\rm gas}$ maximum $\simeq$5550~K. The lower panel shows 
the best fit ({\it thin line}) to the optical spectrum 
around strong $\gamma$-system band-heads of TiO.
\label{lobel:fig3}}
\end{figure*}

\section{Modeling Chromospheric Dynamics of Red Supergiant Betelgeuse}

1-D models of the thermodynamic and dynamic structure of $\alpha$ Ori's 
photosphere and inner chromosphere (to $\sim$10 $\rm R_{*}$) are developed 
in \citet{lob00}. Modern methods in radiative transfer modeling 
venture well beyond outdated approaches that, for example, can 
only consider escape probabilities and radio fluxes. 
The latter flux-integrated models completely fail to 
produce fundamental chromospheric lines (e.g., Balmer H, and 
Mg~{\sc ii}, Ca~{\sc ii}, K~{\sc i} resonance lines, etc.). 
Accurate models (1-D or multi-D) for these important spectral lines 
are required for unraveling the wind-driving mechanism and related 
heating physics in the outer atmospheres of cool stars. 
The transport of radiation does 
not discriminate between so-called `warm and cool gas components'. 
In transfer calculations (irrespective of the number of dimensions), 
the model temperature and electron density structures ($N_{\rm e}$) 
must properly excite chromospheric absorption (e.g., H$\alpha$, H$\beta$) 
and emission lines, including the central self-reversals of optically 
thick emission lines. The model in Fig. 3 shows the radial mean 
values of $T_{\rm gas}$ and $N_{\rm e}$ up to $\sim$10 $\rm R_{*}$, based on 
detailed fits to the classical chromospheric indicators observed 
in $\alpha$ Ori. It correctly produces the observed 
near-UV emission line spectrum. A model without a gas kinetic temperature 
structure increasing to $\sim$5550 K in the inner chromosphere 
cannot excite the subordinate H$\alpha$-transition of $\sim$300 m\AA. 
The entire absorption line would become invisible with respect to 
strong TiO absorption in the cooler photosphere \citep{lob00}. 
The Mg~{\sc ii} $h$ \& $k$ emission lines would also vanish. The fully 
intensity-saturated self-scattered cores of these broad lines 
(about the strongest chromospheric emission lines observed in any 
cool star) result from a very large column of (warm) chromospheric 
plasma that dominates the overall gas kinetic temperature- and 
density-structure of the inner chromosphere. A much steeper 
outwardly decrease of $N_{\rm e}$ (i.e., by an order of magnitude over 2 $\rm R_{*}$) 
yields too large decrements of total widths computed for spatially resolved 
chromospheric emission lines. Large variations of chromospheric 
emission line widths are consistently not observed in spatially resolved 
raster scans across the inner chromosphere with HST-$STIS$ in 1998$-$1999. 
The macro-broadening velocity is supersonic (9$\pm$1 $\rm km\,s^{-1}$), 
and observed to be highly isotropic and homogeneous \citep{lob01} 
across the inner chromospheric disk.

\begin{figure*}[!t]
\plottwo{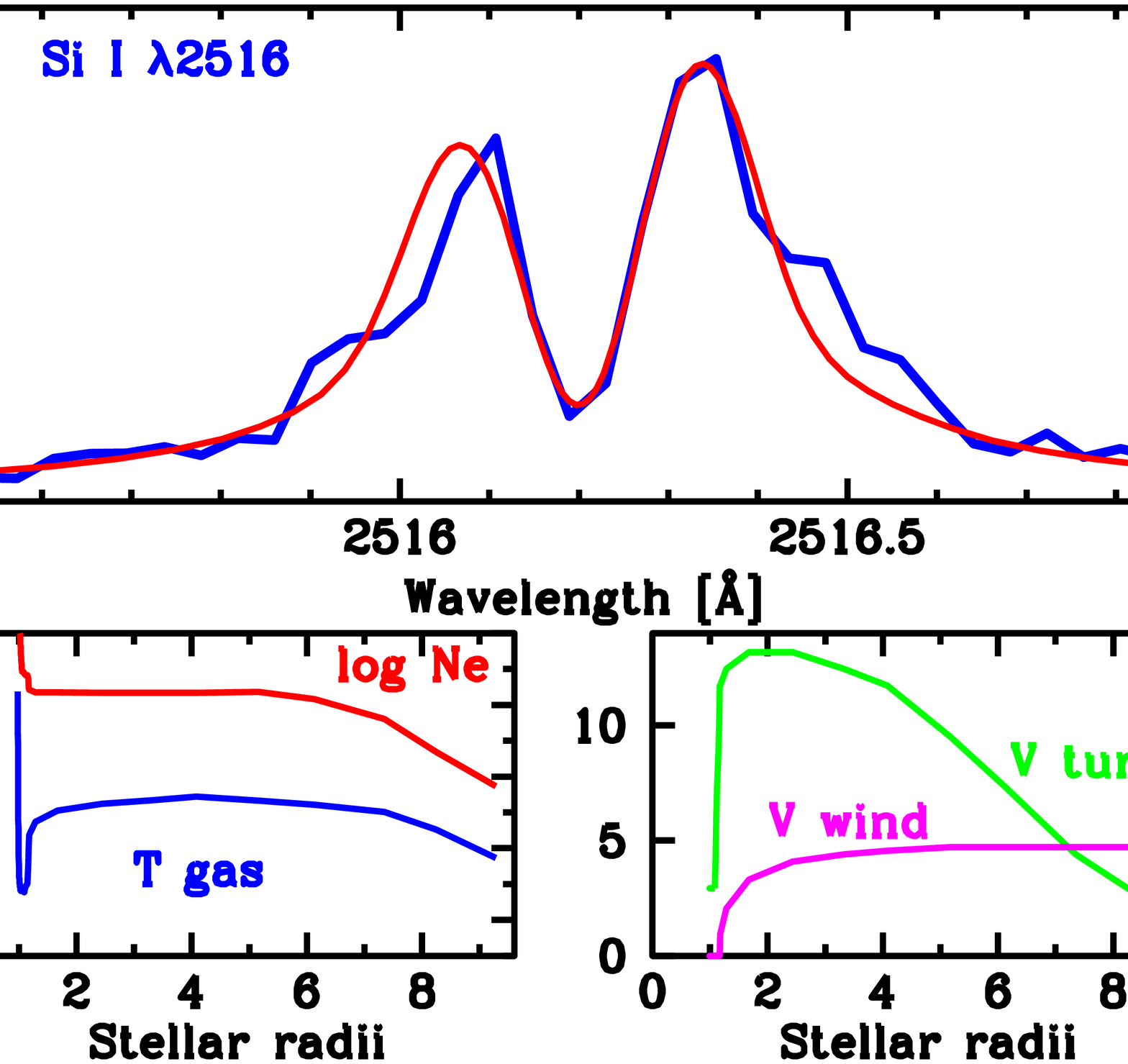}{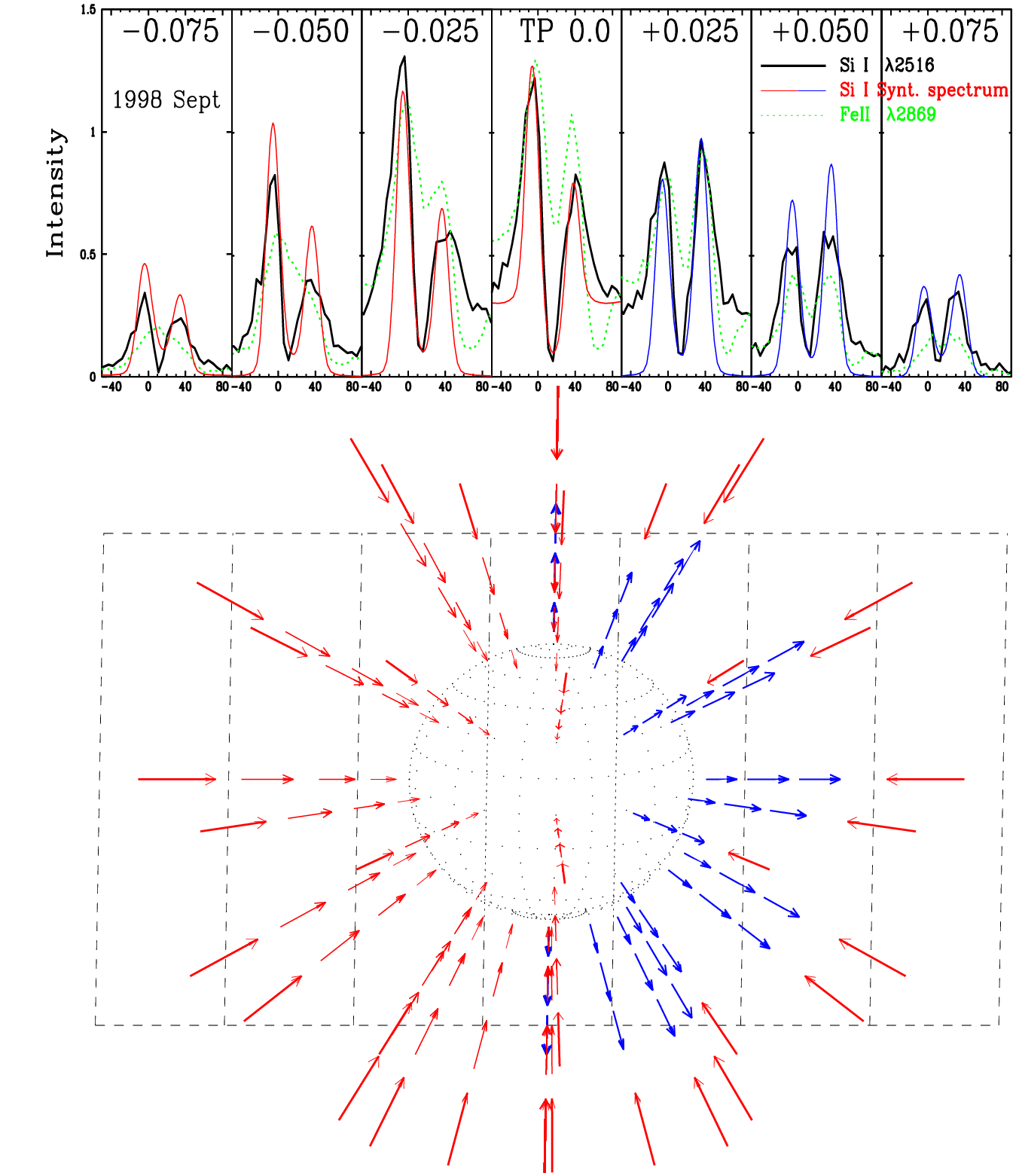} 
\caption[]{{\it Panels left:} Detailed non-LTE fit 
({\it thin line in upper panel}) to the spatially resolved 
Si~{\sc i} $\lambda$2516 emission line observed in $\alpha$ Ori 
at 75 mas with HST-$STIS$ ({\it boldly drawn line}). The 
radial structures of the model 
are shown in the lower panels. $T_{\rm gas}$ is in kK units, 
and Vwind and Vturb in $\rm km\,s^{-1}$. {\it Panels right:} 
3-D mean flow velocities are inferred from detailed fits 
({\it thin drawn lines in upper panel}) to the asymmetric shape 
of the Si~{\sc i} line ({\it bold lines}). The reversal of 
the emission line component intensity reveals simultaneous 
up- and downflow across the front hemisphere in 1998 September.
\label{lobel:fig4}}
\end{figure*}

Detailed transfer fits to the temporal variability and 
spatial intensity reversals observed in the emission line 
components of Si~{\sc i} $\lambda$2516 
reveal that the inner chromosphere oscillates asymmetrically with 
simultaneous up-and downflow ($\pm$4 $\rm km\,s^{-1}$) 
across the front hemisphere ({\it Fig. 4}). Interestingly, 
\citet{fre08} recently presented 3-D hydrodynamic 
models with large asymmetric movements of the photosphere and 
strong ($M_{\rm rad}$$>$2) radially propagating shock 
waves. Strong shock waves are an efficient mechanism for transporting 
mechanical momentum very far from the stellar surface. 
They can support the co-existence of dust-sustaining 
low-temperature conditions ($T_{\rm dust}$$\simeq$600 K) and 
(possibly shock-excited) chromospheric emission lines observed with $STIS$ 
far inside the circumstellar dust envelope \citep[{{\it Fig. 5}};][]{lob05} 
up to 3$\arcsec$ ($>$100 $\rm R_{*}$).

\begin{figure*}[!t]
\plottwo{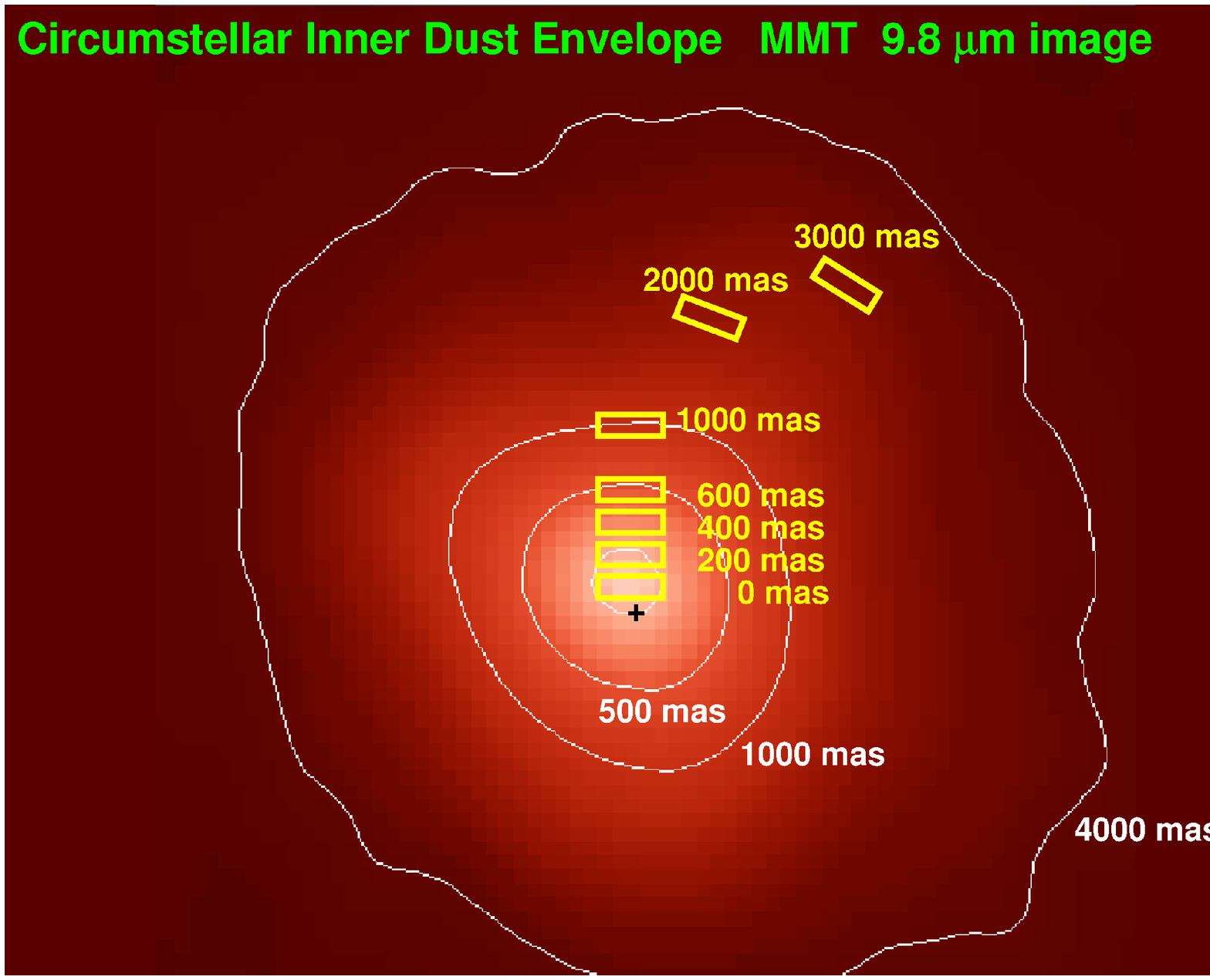}{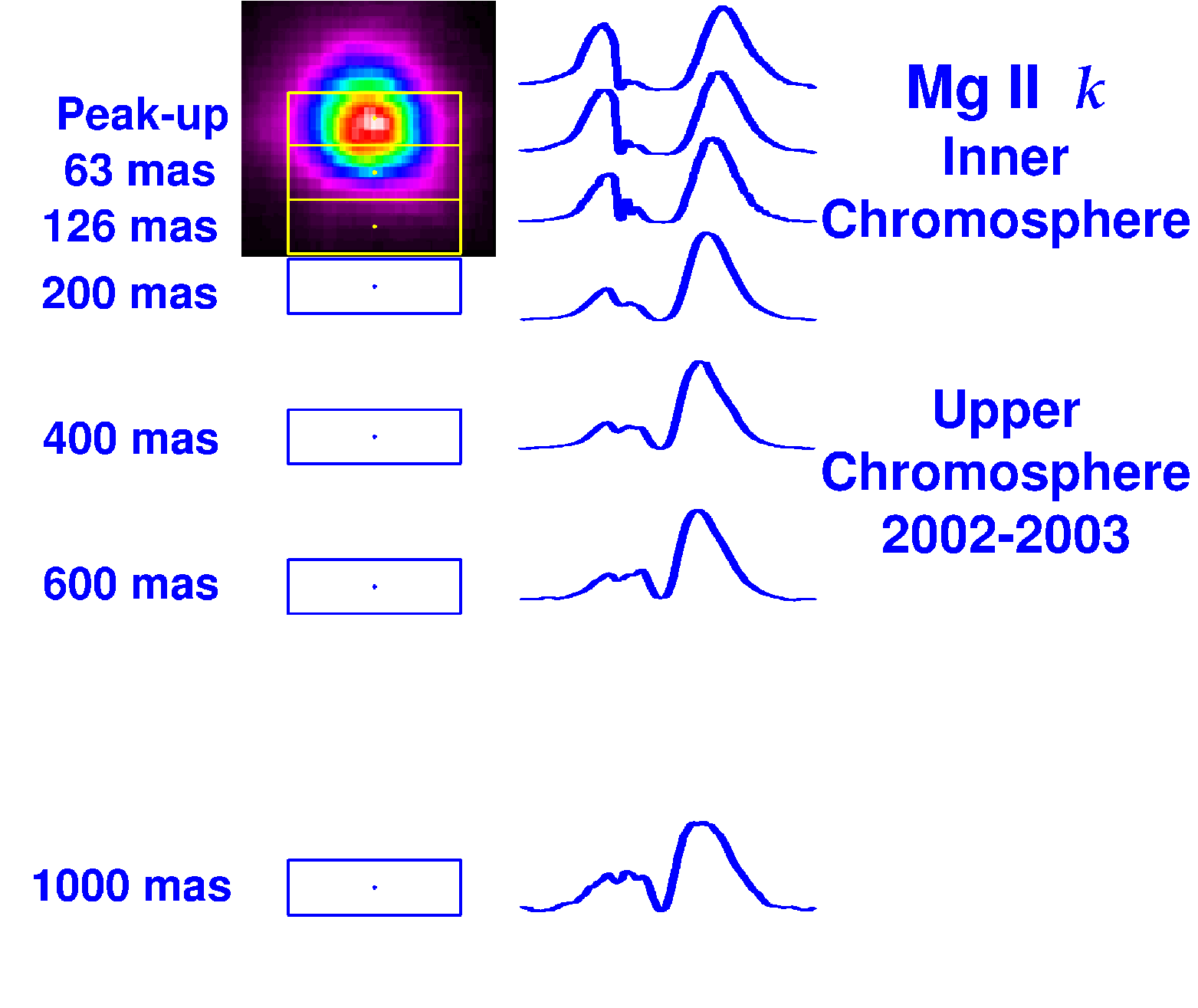} 
\caption[]{{\it Panel left:} Relative aperture positions of spatially 
resolved spectroscopic observations with HST-$STIS$ in 2002-2003 
compared to a 9.8 $\mu$m image (MMT-$MIRAC2$) of $\alpha$ Ori's inner 
circumstellar dust envelope \citep[{\it adapted from}][]{hin98}. 
{\it Panel right:} The Mg~{\sc ii} $k$ resonance emission line observed 
across the inner and upper chromosphere with the 63$\times$200 mas 
aperture compared to a false color near-UV image (HST-$FOC$). 
The line is scaled to equal intensities and reveals important 
profile shape changes of the central self-scattering core, 
signaling outward acceleration of the chromospheric wind.
\label{lobel:fig5}}
\end{figure*}

\section{Conclusions}

Mechanical wave action (provided by circumstellar density waves,  
propagation of pressure waves, and astrophysical turbulence) 
due to atmospheric pulsation is a very important aspect 
for studying the detailed physics of active structuring, 
acceleration, and heating mechanisms of the extended winds 
and circumstellar environments of hot and cool massive stars. 
Recent advances in quantitative spectroscopy that combine 
3-D radiative transfer methods with multi-D hydrodynamic 
and semi-empiric modeling provide powerful new methods for 
pinpointing these mechanisms.

\acknowledgements 
The contributions from many $\rho$ Cas observers over the past 15 years
are gratefully acknowledged. This research is based in part on data of 
$\alpha$ Ori obtained with the NASA/ESA Hubble Space Telescope, 
collected at the STScI, operated by AURA Inc., under contract NAS5-26555. 
Financial support has been provided by STScI grant HST-GO-09369.01. 
The Belgian Federal Science Policy is gratefully acknowledged 
for financial support for 3-D radiative transfer modeling research
of massive stars.

\end{document}